# Hydrogen storage properties of Mn and Cu for Fe substitution in TiFe$_{0.9}$ intermetallic compound

Erika M. Dematteis,[a,b)*] Fermin Cuevas,[a)] and Michel Latroche[a)]

[a)] Univ Paris Est Creteil, CNRS, ICMPE, 2 rue Henri Dunant, 94320 Thiais, France

[b)] Present address: Department of Chemistry, Inter-departmental Center Nanostructured Interfaces and Surfaces (NIS), and INSTM, University of Turin, Via Pietro Giuria 7, 10125 Torino, Italy

*Corresponding author: Erika M. Dematteis

E-mail address: erikamichela.dematteis@unito.it

**Abstract**

The present study investigates the partial substitutions of Mn and Cu for Fe in the TiFe-system to gain better understanding of the role of elemental substitution on its hydrogen storage properties. The TiFe$_{0.88-x}$Mn$_{0.02}$Cu$_x$ ($x$ = 0, 0.02, 0.04) compositions were studied. From X-Ray Diffraction (XRD) and Electron Probe Micro-Analysis (EPMA), it was found that all alloys are multi-phase, with TiFe as a major phase, together with β-Ti and Ti$_4$Fe$_2$O-type as secondary precipitates, of all them containing also Mn and Cu. Increasing the Cu content augments the secondary phase amounts. Low quantity of secondary phases helps the activation of the main TiFe phase for the first hydrogen absorption, but on increasing their amounts, harsher activation occurs. Both Mn and Cu substitutions increase the cell parameter of TiFe, thus decreasing the first plateau pressure. However, Cu substitution rises the second plateau pressure revealing the predominancy of electronic effects associated to this substitution. All samples have fast kinetics and high hydrogen capacity making these substituted compounds promising for large scale stationary applications.

# Introduction

The increase in air pollution and global temperatures demonstrate how essential it is to look for alternatives to fossil fuels. Renewable energy will enable the transition towards $CO_2$-free energy, but its intermittent nature will induce needs for high capacity storages. Hydrogen can be produced from renewable sources and, as an efficient energy carrier, can be stored for a long period. Metal hydrides are safe materials for solid-state hydrogen storage under mild conditions. TiFe is a low-cost and efficient *AB*-type intermetallic compound for hydrogen storage. It crystallizes in the CsCl-type cubic structure with Ti and Fe occupying the 1*a* (0,0,0) and 1*b* (½,½,½) sites respectively, in *Pm*-3*m* space group. Upon hydrogen absorption, the consecutive formation of the monohydride, β-TiFeH, and dihydride, γ-TiFeH$_2$, occurs, with a total volume expansion of 18% and a maximum gravimetric capacity of 1.86 wt.% $H_2$.[1] However, TiFe exhibits drawbacks and particular features as hydrogen storage material. Firstly, it is difficult to activate towards the first hydrogen sorption. Secondly, Pressure-Composition-Isotherms (PCI) are characterized by two subsequent plateau pressures. All these properties are at the origin of controversial results concerning the crystal structure of TiFe hydrides/deuterides.[1]

Considering the Ti-Fe phase diagram, the TiFe intermetallic exists in a range of composition, from 49.7 to 52.5 Ti at%.[2,3] The 1:1 stoichiometry presents harsh activation temperatures (up to 400°C), and, at 30°C, the second plateau occurs at 2 MPa, which is rather high for practical applications.[4,5] Therefore, there always have been the interest in improving activation properties and lowering or tailoring the plateau pressure by elemental substitution to tune the hydrogenation properties of TiFe-based compounds as a function of final targets. Until recently, studies have been conducted to investigate element additions such as yttrium[6] or hafnium[7] to the TiFe system showing that Hf is slightly soluble in TiFe, lowering the equilibrium pressure and improving activation, whereas Y does not seem to dilute in the main phase but forms precipitates that enhance both activation and kinetics. Elemental substitution in TiFe by ball milling has been recently investigated for Co and Nb, evidencing a beneficial effect in flattering plateau in PCI curves, enhancing resistance to poisoning and upon cycling, even if a drastic reduction of the capacity was caused by the suppression of the β to γ transition.[8]

Off-stoichiometry, respect to 1:1, and/or elemental substitutions in TiFe can change significantly both activation processes and hydrogen storage properties. Interestingly, it has been shown that the off-stoichiometric compound TiFe$_{0.9}$, or more correctly Ti$_{1a}$(Fe$_{0.947}$Ti$_{0.053}$)$_{1b}$ as partial Ti substitution for Fe occurs in the 1$b$ site of the cubic CsCl structure, requires almost no activation process for the first hydrogenation.[9] Partial substitution of Fe by Mn is also reported to reduce the need of alloy activation and, moreover, promotes lower equilibrium pressures at room temperature.[10] Thus, substituted Ti(Fe$_{1-x}$Mn$_x$)$_{0.9}$ alloys combine easy activation and low plateau pressures at room temperature, being good candidates for hydrogen storage applications near normal conditions of pressure and temperature. Recently, Ali *et al.* investigated the simultaneous substitution of Y and Cu in TiFe$_{1-x}$Mn$_x$ alloys, evidencing a further reduced plateau pressure.[11,12] Yttrium substitutes Ti promoting capacity and lowering hysteresis, whereas Cu substitution for Fe promotes easier activation and faster kinetics. Another example of quaternary tailoring that remarkably improved activation has been recently investigated by Leng *et al.*, where the addition of Ce to TiFe$_{0.90}$Mn$_{0.10}$ evidenced no change in thermodynamics but an improvements of kinetics with a modest decrease in hydrogen capacity.[13]

Back in 1986, Nagai *et al.* investigated the hydrogen storage properties of TiFe$_{1-x}$Cu$_x$ alloys mixed with Fe$_2$O$_3$. Short incubation time for activation was reached thanks to the presence of Fe$_2$O$_3$ and secondary phases.[14] Wu *et al.* investigated in detail the crystal structure of TiFe$_{1-x}$Cu$_x$ alloys showing that the cell parameter of the cubic phase linearly increases with the Cu amount.[15]

As a matter of fact, chemical substitution in TiFe, as previously shown for Mn and Cu, can be used to tune plateau pressures to the sorption conditions required by the final application. For this reason, in this work, characteristics of Mn and Cu-substituted TiFe alloys will be discussed, featuring their microstructural and hydrogen storage properties, *i.e.* both thermodynamics and kinetics, and activation issues.

## Experimental

**Sample preparation**

All TiFe-based alloys (approx. 8 g) were prepared by induction melting of bulk elements under argon in a water-cooled copper crucible. The pure elements were cleaned from oxides at the surface by hand polishing and weighted to the wanted compositional ratio. A primary alloy containing only Ti (Alfa Aesar, 99.99%) and Fe (Alfa Aesar, 99.97%) was melted firstly. Secondly Mn (Goodfellow, 99.98%) and Cu (Alfa Aesar, 99.999%) were added, and the ingots were turned over and melted three times to enhance homogeneity. The samples, wrapped in a tantalum foil to prevent contamination, were introduced into a silica tube. The tube was sealed under argon atmosphere, annealed in a resistive furnace at 1000°C for 1 week, and quenched into water to room temperature. Weight losses lower than 0.1% were detected during melting and annealing. After synthesis and annealing, the ingots were crushed under air into millimetre size pieces for metallographic examination. Then the samples were introduced in an Ar purified glovebox and smashed into fine powder down to sub-millimetre size. Argon-filled glove boxes with a circulation purifier were used to store, prepare and manipulate the powders. $O_2$ and $H_2O$ levels were lower than 1 ppm to avoid oxidation and degradation of the material.

**Characterizations**

*Electron Probe Micro-Analysis (EPMA)*

Metallographic examination and elemental analysis by electron probe micro-analysis (Cameca SX100) were performed to check the homogeneity and phase composition of the alloys. Precipitation of minor phases in the TiFe matrix was detected in Back-Scattered Electron (BSE) images. Acceleration voltage and beam current were 15 kV and 40 mA, respectively. The composition of the different phases was determined by Wavelength-Dispersive X-ray Spectroscopy from 10 to 100 point measurements at the surface of the polished samples. The standard deviation of these measurements is given as the uncertainty of the phase composition. This uncertainty results from counting statistics, experimental setup, size of precipitates and local chemical fluctuations.

*Powder X-ray diffraction (PXD)*

Power X-Ray Diffraction patterns were obtained at room temperature on a Bruker D8 Advance Bragg Brentano diffractometer using Cu-Kα radiation (λ=1.5418 Å), in 2θ-range of 20–120°, step size 0.02, time step 7 seconds. All patterns were refined by the Rietveld method[16] using FullProf package.[17]

*Hydrogenation properties*

Activation, kinetic studies and Pressure Composition Isotherm (PCI) curves were monitored with a home-made Sieverts' type apparatus. Approx. 500 mg of sample were loaded into stainless steel sample holder inside the glove box under argon atmosphere, then connected to the rig, evacuated under primary vacuum and activated by exposing them to pure gaseous hydrogen (scientific hydrogen 6.0 $H_2$ from Linde) under different pressure and temperature conditions depending on the investigated alloy.

PCI curves were measured after five hydrogen absorption-desorption cycles to ensure full activation and reproducible PCI isotherms. An oil or water thermostatic bath was used at the selected temperature (25, 40 and 55°C) to maintain the temperature during PCI recording. PCI curves at different temperatures were measured by the manometric Sieverts' method. Prior each PCI, the sample was dehydrogenated at 200°C under primary vacuum for 30 minutes.

## Results and Discussion

To investigate the role of Mn and Cu substitutions, three different compositions have been synthetized: TiFe$_{(0.88-x)}$Mn$_{0.02}$Cu$_x$ with $x = 0$, 0.02, 0.04. For the sake of simplicity, these alloys will be hereafter named as Cu0, Cu2 and Cu4. Since a previous study evidenced that 2 at% Mn introduced in the alloy lowered the average desorption plateau pressure close to 0.1 MPa at room temperature,[10] a low amount of Mn (~1 at%) was fixed, while the amount of Cu in the alloy was variated (**Table 1**).

**Table 1** – Elemental analysis (in at.%) of phases as determined by EPMA for Cu0, Cu2 and Cu4 samples. Data reported with no error comes from single point analysis of secondary phases due to limited spatial distribution.

| Sample – nominal composition | Phase | Ti (at%) | Fe (at%) | Mn (at%) | Cu (at%) |
|---|---|---|---|---|---|
| **Cu0** – $Ti_{52.6}Fe_{46.3}Mn_{1.1}$ | TiFe | 51.9±0.1 | 47.2±0.2 | 0.94±0.07 | 0 |
| | $Ti_4Fe_2O$ | 67 | 32 | 1 | 0 |
| | β-Ti | 78 | 21 | 1 | 0 |
| **Cu2** – $Ti_{52.6}Fe_{45.2}Mn_{1.1}Cu_{1.1}$ | TiFe | 51.2±0.2 | 46.8±0.3 | 0.88±0.12 | 1.06±0.07 |
| | $Ti_4Fe_2O$ | 62 | 36 | 1 | 1 |
| | β-Ti | 78.4±0.3 | 19.7±0.3 | 1.12±0.02 | 0.81±0.01 |
| **Cu4** - $Ti_{52.6}Fe_{44.2}Mn_{1.1}Cu_{2.1}$ | TiFe | 51.8±0.1 | 45.1±0.3 | 1.03±0.09 | 2.10±0.13 |
| | $Ti_4Fe_2O$ | 67 | 30 | 1 | 2 |
| | β-Ti | 78 | 20 | 1 | 1 |

*Chemical, structural and morphological characterization*

**Figure 1** shows the metallographic examination of the three samples. All microstructures can be described as consisting of a major TiFe phase (matrix) and different amounts of secondary phases. Cu0 presents small quantity of isolated precipitates while Cu2 and Cu4 have similar microstructures with precipitation of secondary phases at the grain boundaries. The highest amount of secondary phase is clearly observed for Cu4 (**Figure 1, c**). This result is in good agreement with literature studies that report the reduction of the homogeneity range of TiFe due to Mn and Cu substitution.[18,19]

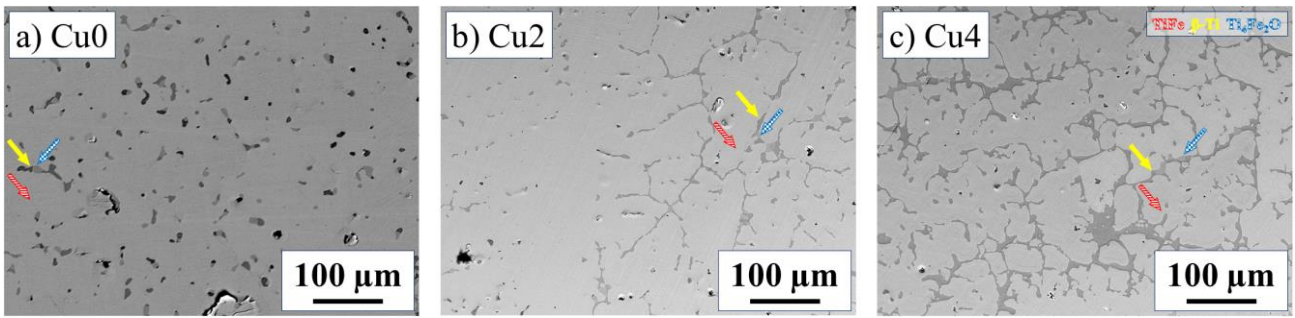

**Figure 1** – BSE metallographic images of a) Cu0 - TiFe$_{0.88}$Mn$_{0.02}$, b) Cu2 - TiFe$_{0.86}$Mn$_{0.02}$Cu$_{0.02}$ and c) Cu4 - TiFe$_{0.84}$Mn$_{0.02}$Cu$_{0.04}$. Representative regions for different phases are pointed by arrows in the images.

Results of EPMA analysis for the different phases observed in the three samples are reported in **Table 1**. The low standard deviation for the atomic content in the major matrix phase evidences a homogenous composition of the TiFe phase in all samples, thanks to the performed annealing. For secondary phases, their composition could be well determined (standard deviation given) when the precipitate size is above the spatial resolution of EPMA analysis (~ 1μm$^3$). The matrix can be assigned to TiFe intermetallic compound containing Mn and Cu (when present in the alloy). In these samples, because of the off-stoichiometry respect to the 1:1 Ti-Fe composition, and nominal composition close to the Ti-rich limit of homogeneity region, the TiFe matrix has a Ti content of approx. 52 at%, and the β-Ti solid solution phase has a Fe content of approx. 20 at.% in accordance with the Ti-Fe binary phase diagram.[3] A second precipitate with Ti:Fe ratio of 2:1 is detected as well (**Table 1**), which is assigned to the formation of Ti$_4$Fe$_2$O as will be later confirmed by XRD. Generally, Mn and Cu are homogenously distributed in all phases. In fact, as it can be observed from **Table 1**, EMPA analysis evidenced that, not only the TiFe phase, but also secondary phases contain Mn and Cu in amounts proportional to the nominal composition of the sample.

Due to the limited amount of oxide in the sample, this phase is difficult to detect by SEM (**Figure 1**). Nevertheless, the Ti-rich β-Ti phase can be observed as dark grey zone in **Figure 1**, while Ti$_4$Fe$_2$O, that has a higher amount of Fe, can be observed as light grey zone. Furthermore, EMPA point analysis evidenced that the oxide phase is localised at the grain boundaries, close to the β-Ti phase.

XRD patterns for the three samples are displayed in **Figure 2**. Diffraction peaks can be indexed with three phases in agreement with EPMA analysis. The major phase is TiFe-based adopting CsCl-type structure. Additional low intensity peaks are indexed with two secondary phases: β-Ti and $Ti_4Fe_2O$. β-Ti is the high temperature *bcc* allotropic form of titanium. It has an iron solubility limit of *ca.* 22 at.% at 1085 °C as reported in the binary Ti-Fe phase diagram.[3] $Ti_4Fe_2O$ has a cubic structure as well (S.G. *Fd-3m)*, with 2:1 Ti:Fe ratio, in agreement with EPMA analysis. This ternary compound is stabilized by oxygen and is reported in the Ti-Fe-O phase diagram.[20] It should be noted that the oxygen-free $Ti_2Fe$ phase does not exist in the binary Ti-Fe phase diagram.

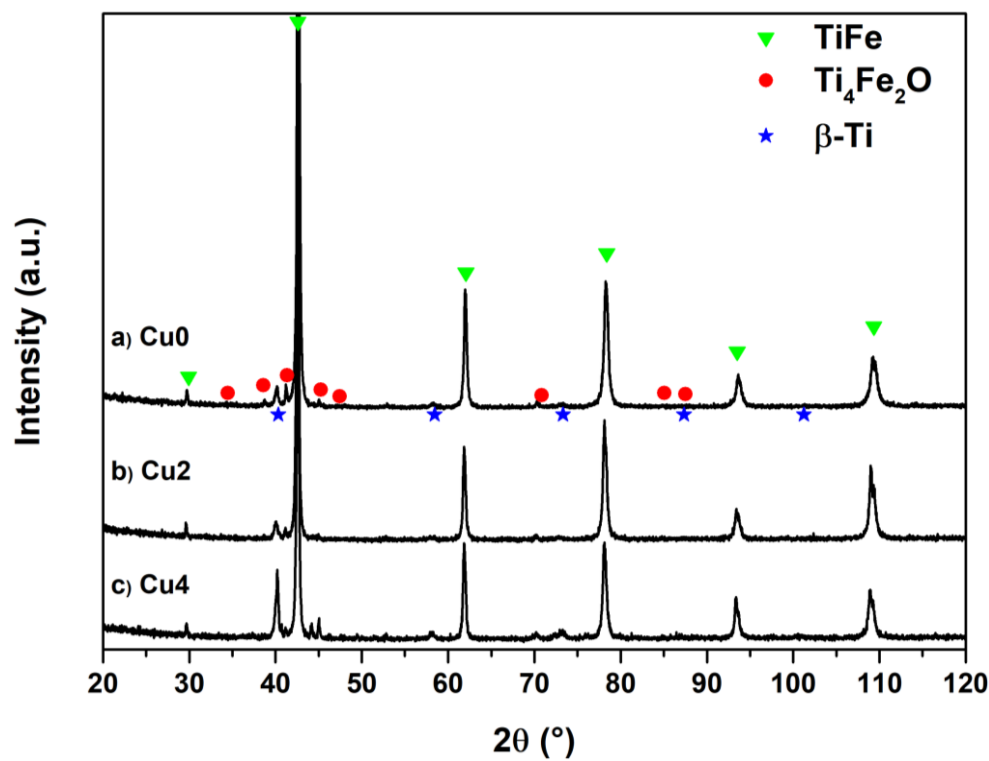

**Figure 2** – X-Ray diffraction patterns of the synthetized alloys after annealing at 1000°C for 1 week. Detailed data for the phases identified in the patterns are given in **Table 1** and **Table 2**.

**Table 2** shows phase abundances determined by Rietveld refinement (**Figure S1**). Clearly, the amount of β-Ti solid solution increases with the Cu content. This is attributed to the fact that the narrow homogeneity region of binary TiFe, which ranges from 49.7 to 52.5 at.% Ti, gradually shrinks with increasing Cu substitution as displayed in the ternary Cu-Fe-Ti phase diagram.[21] On the other hand, $Ti_4Fe_2O$ amount does not vary continuously with $x_{Cu}$. Compared to the Cu-free sample Cu0, it slightly decreases for Cu2 and then increases for higher Cu amount Cu4. Likely, the oxide phase does not depend on Cu content but rather on raw material impurities or residual oxygen during alloy synthesis. The lattice parameter of the major TiFe phase tends to increase with the amount of Mn, and further increases introducing Cu (**Table 2**). The lattice parameter of Cu-free sample (Cu0, $TiFe_{0.88}Mn_{0.02}$, 1 at% Mn) is equal to the one of $TiFe_{0.85}Mn_{0.05}$ reported in the literature,[10] where a higher amount of Mn was introduced in the alloy (2 at% Mn).

**Table 2** – Phase abundances and lattice parameter of TiFe phase for the three alloys here synthetized as determined by Rietveld analysis of XRD data. Results are compared with literature data for $TiFe_{0.90}$[9] and $TiFe_{0.85}Mn_{0.05}$[10].

| Sample | Nominal composition | TiFe a (Å) | TiFe (wt.%) | β-Ti (wt.%) | $Ti_4Fe_2O$ (wt.%) |
|---|---|---|---|---|---|
| **[Guéguen2011][9]** | $TiFe_{0.90}$ | 2.981(9) | 94.8 | 2.1 | |
| **[Challet2005][10]** | $TiFe_{0.85}Mn_{0.05}$ | 2.985 | 97 | | |
| **Cu0** | $TiFe_{0.88}Mn_{0.02}$ | 2.985(8) | 95.0±0.5 | 2.6±0.5 | 2.3±0.5 |
| **Cu2** | $TiFe_{0.86}Mn_{0.02}Cu_{0.02}$ | 2.988(2) | 94.9±0.6 | 3.5±0.5 | 1.6±0.5 |
| **Cu4** | $TiFe_{0.84}Mn_{0.02}Cu_{0.04}$ | 2.991(6) | 86.5±0.7 | 11.0±0.5 | 2.5±0.5 |

*Hydrogenation properties*

*Activation behaviour and kinetics*

All alloys can be activated at room temperature (RT) under 2.5 MPa of $H_2$, except for Cu4 which required 100°C and 5 MPa $H_2$. Under these thermodynamic conditions, the incubation time for the first hydrogen absorption took some tens of hours as displayed in **Table 3** and in pressure as a function of time curves shown in **Figure S2**. The secondary phases β-Ti and $Ti_4Fe_2O$ facilitate the activation process.[20] The introduction of Cu implies either longer incubation time or harsher pressure and temperature conditions of activation. A limit for the secondary-phase amount or Cu content to obtain optimal activation features cannot be defined precisely. However it can be stated that a low amount of secondary phases (around 2-5 wt%) can be beneficial to the activation process, if hydrogen can diffuse easily through them or if they can be hydrogenated under moderate temperature and pressure condition.[20] As a matter of fact, this study shows that by increasing Cu, an increase in secondary phases' amount is evidenced and we suggest that a high amount of β-Ti can hinder or slow down the diffusion of hydrogen during the first hydrogenation of the material. We hypothesize, in agreement with previous works [22], that the key parameter for the alloy activation is the density of interfaces between TiFe and secondary phases. As shown in **Figure 1**, the interface density does not increase with Cu content, since the secondary phases' precipitates are larger but not more numerous (i.e. **Figure 1** c, Cu4). Furthermore, from compositional analysis (**Table 1**), secondary phases contain Cu, which might be less active in dissociating $H_2$ molecules than Cu-free phases.

After activation, the hydrogen uptake at $P_{H_2}$ = 2.5 MPa and 25°C displays fast kinetics with 90% of the full reaction ($t_{90}$) reached in approx. 1 min (**Table 3** and **Figure S2**). The kinetics is slightly improved by introducing Cu in the system.

**Table 3** – Data for the activation ($T$, $P_{H_2}$, incubation time) of Cu0, Cu2 and Cu4 samples, their kinetics ($t_{90}$) in the fifth hydrogenation cycle prior to PCI measurement at 25°C under applied pressure of $P_{H_2}$ = 2.5 MPa, and hysteresis for the first plateau at 25 °C ($ln\frac{P_{abs}}{P_{des}}$).

| Sample | T (°C) | $P_{H_2}$ (MPa) | Incubation time(h) | $t_{90}$ (s) | $ln\frac{P_{abs}}{P_{des}}$ first plateau |
|---|---|---|---|---|---|
| **Cu0** | RT | 2.5 | 6.5 | 69 | 0.68 |
| **Cu2** | RT | 2.5 | 41.5 | 59 | 0.60 |
| **Cu4** | 100 | 5.0 | 15.0 | 46 | 0.42 |

*Thermodynamics*

**Figure 3** shows the PCI curves for all synthetized samples at 25, 40 and 55°C. With the increase of Cu, several features are observed: *i*) decrease of the maximum storage capacity, *ii*) lowering of the first plateau pressure and of its hysteresis, and *iii*) increase of the second plateau pressure leading to a concomitant increase of the pressure difference between the first and the second plateau.

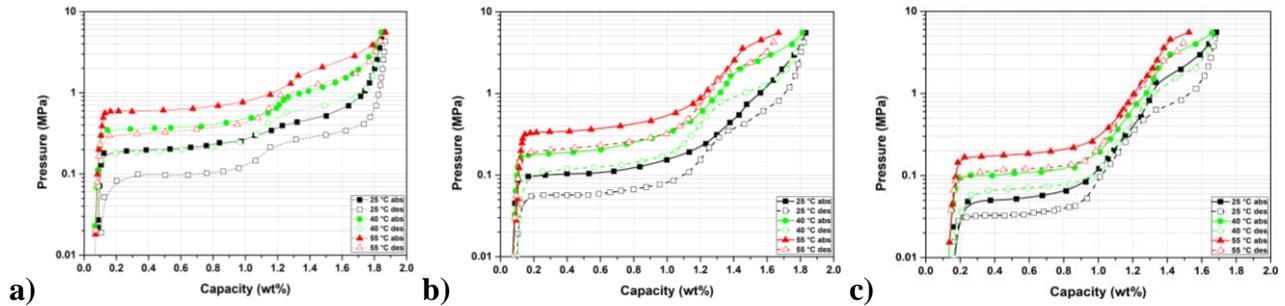

**Figure 3** – Absorption (closed points) and desorption (open points) PCI curves for a) Cu0, b) Cu2 and c) Cu4 samples; at 25 (black squares), 40 (green circles) and 55°C (red triangles). Solid lines in absorption and dashed lines in desorption are a guide for the eyes.

The maximum capacity at $P_{H_2}$ = 5.5 MPa and 25°C gradually decreases with Cu content: 1.86, 1.83 and 1.68 wt% for Cu0, Cu2 and Cu4, respectively. In addition, at low pressure, $P_{H_2}$ = 0.02 MPa, hydrogen trapping tends to increase with Cu content, being equal to 0.09 wt% for Cu0 and Cu2 samples and 0.16 wt% for Cu4. Hydrogen absorption at low pressure is attributed to hydrogen trapping both in α-TiFe and in β-Ti. The α-solid solution in TiFe extends up to TiFeH$_{0.1}$ (~ 0.09 wt%) which would account for hydrogen trapping in samples Cu0 and Cu2. Additional trapping for sample Cu4 is attributed to β-Ti, the amount of which is as high as 11 wt% for this sample as displayed in **Table 2**. Thus, Cu4 has the lowest reversible and maximum capacity since it contains the highest amount of β-Ti. We argue that during alloy activation, a TiH$_2$-type phase is formed that partially desorbs on heating at 200°C before PCI experiments. During PCI, a hydrogen pressure as low as 0.02 MPa is enough to rehydrogenate this phase, contributing to trapping phenomena and reforming titanium dihydride, which remains stable at the PCI thermodynamic conditions. To verify the latter hypothesis, XRD and Rietveld refinement analysis were conducted on dehydrogenated samples after PCI experiments (**Figure S3** and **Table S1**). The occurrence and stability of the $TiH_2$-type phase, which amount increases with Cu-content, is corroborated. Moreover, the reversible capacity of Cu0, Cu2 and Cu4 samples at 25°C for a pressure window comprised between 0.02 and 5.5 MPa gradually decreases following the sequence 1.75, 1.74 and 1.51 wt%, respectively. This decrease confirms the hydrogen stored in the β-Ti solid solution is not reversible at the studied temperatures in agreement with Matsumoto *et al*.[23]. Recently, the interaction between hydrogen and β-Ti for the same composition as in the present work, $Ti_{80}Fe_{20}$, was studied by Fokin *et al.*,[24] following a previous report of Verbetskij *et al.*[25]. At room temperature a highly stable hydride phase is formed with the composition $Ti_{0.8}Fe_{0.2}H_{\sim1.7}$, adopting the structure of titanium hydride.[24]

Concerning the first plateau pressure, the Cu0 sample, with nominal composition $TiFe_{0.88}Mn_{0.02}$, has the highest value among all samples here synthetized and is comparable with the data reported for $TiFe_{0.85}Mn_{0.05}$ (**Figure S4**).[10] Thus, a Mn content as low as 1 at% significantly modifies the hydrogen storage properties of the alloy as compared to $TiFe_{0.90}$.[9] Moreover, when 1 at% of Cu is additionally substituted into $TiFe_{0.88}Mn_{0.02}$ (*i.e.* $TiFe_{0.86}Mn_{0.02}Cu_{0.02}$ for Cu2 sample), the equilibrium pressure of the first plateau further decreases (**Table 4**). This tendency is confirmed for Cu4 sample ($TiFe_{0.84}Mn_{0.02}Cu_{0.04}$) which presents an even lower equilibrium pressure for the first plateau.

It is worth comparing the lowering of the first plateau pressure for both Mn and Cu substitutions (**Table 4**) with the crystal data of the TiFe-based intermetallic compounds (**Table 2**). Both substitutions enlarge the lattice parameter of the TiFe phase. If one assumes that both Mn and Cu replace Fe in the TiFe crystal structure, the observed cell volume expansion correlates with the smaller atomic radii of Fe, $r_{Fe}$ = 1.24 Å, as compared to Cu, $r_{Cu}$ = 1.28 Å, and Mn, $r_{Mn}$ = 1.35 .[26]

Table 4 – Hydrogenation thermodynamics of Cu0, Cu2 and Cu4 samples determined from Van't Hoff plots of PCI curves at different temperatures (25, 40, 55 °C). Plateau pressure data given at hydrogen contents of 0.6 (first plateau) and 1.4 wt.% H (second plateau). Plateau pressure data at 25°C for TiFe$_{0.9}$ are given from ref.[9] for comparison purposes.

| Sample | Plateau | Absorption | | | | | Desorption | | | | |
| --- | --- | --- | --- | --- | --- | --- | --- | --- | --- | --- | --- |
| | | Plateau Pressure (MPa) | | | Enthalpy (kJ/molH$_2$) | Entropy (kJ/molH$_2$ K) | Plateau Pressure (MPa) | | | Enthalpy (kJ/molH$_2$) | Entropy (kJ/molH$_2$ K) |
| | | 25°C | 40°C | 55°C | | | 25°C | 40°C | 55°C | | |
| TiFe$_{0.9}$ | first | 0.337 | | | | | 0.160 | | | | |
| Cu0 | first | 0.205 | 0.374 | 0.614 | -29.8 | -0.106 | 0.099 | 0.200 | 0.337 | 33.3 | 0.112 |
| Cu2 | first | 0.106 | 0.209 | 0.372 | -34.1 | -0.115 | 0.058 | 0.127 | 0.232 | 37.6 | 0.122 |
| Cu4 | first | 0.054 | 0.110 | 0.185 | -33.4 | -0.107 | 0.034 | 0.071 | 0.122 | 34.7 | 0.107 |
| TiFe$_{0.9}$ | second | 0.998 | | | | | 0.471 | | | | |
| Cu0 | second | 0.470 | 1.056 | 1.825 | -27.2* | -0.106* | 0.279 | 0.643 | 1.172 | 30.3* | 0.112* |
| Cu2 | second | 0.482 | 1.776 | 2.492 | -29.3* | -0.115* | 0.372 | 0.902 | 1.970 | 32.5* | 0.122* |
| Cu4 | second | 1.613 | 2.646 | 4.016 | -25.1* | -0.107* | 0.700 | 1.309 | 2.646 | 26.9* | 0.107* |

\* Entropy values for the second plateau fixed to those of the first plateau to mitigate sloping effects

**Figure 4** shows the dependence between plateau pressure and cell volume, showing that the logarithm of the first plateau pressure (in red in **Figure 4**) decreases linearly with the TiFe cell volume. This indicates that hydrogenation thermodynamics of *AB*-type TiFe alloys with Mn and Cu substitutions fulfil well-established empirical geometric models,[27] being the monohydride stabilized when the intermetallic cell volume increases. This geometric rule of hydride stability has been previously observed in many *AB$_5$* and *AB$_2$* alloys.[28–30]

As concerns the second plateau pressure, hydrogenation properties are more intricate (blue triangles in **Figure 4**). With Mn substitution, the plateau pressure decreases, both in absorption and desorption, when comparing data for TiFe$_{0.9}$ and TiFe$_{0.88}$Mn$_{0.02}$ (Cu0) at 25°C (**Table 4**). However, with Cu substitution, the second plateau shifts to higher-pressure values, both in absorption and desorption for all measured temperatures. Now, the logarithm of the plateau pressure increases linearly on decreasing the cell volume. Interestingly, the same trend as that of Cu substitution was observed by Challet *et al*. for Ni substitution in TiFe$_{0.70}$Mn$_{0.20-x}$Ni$_x$ alloys.[10]

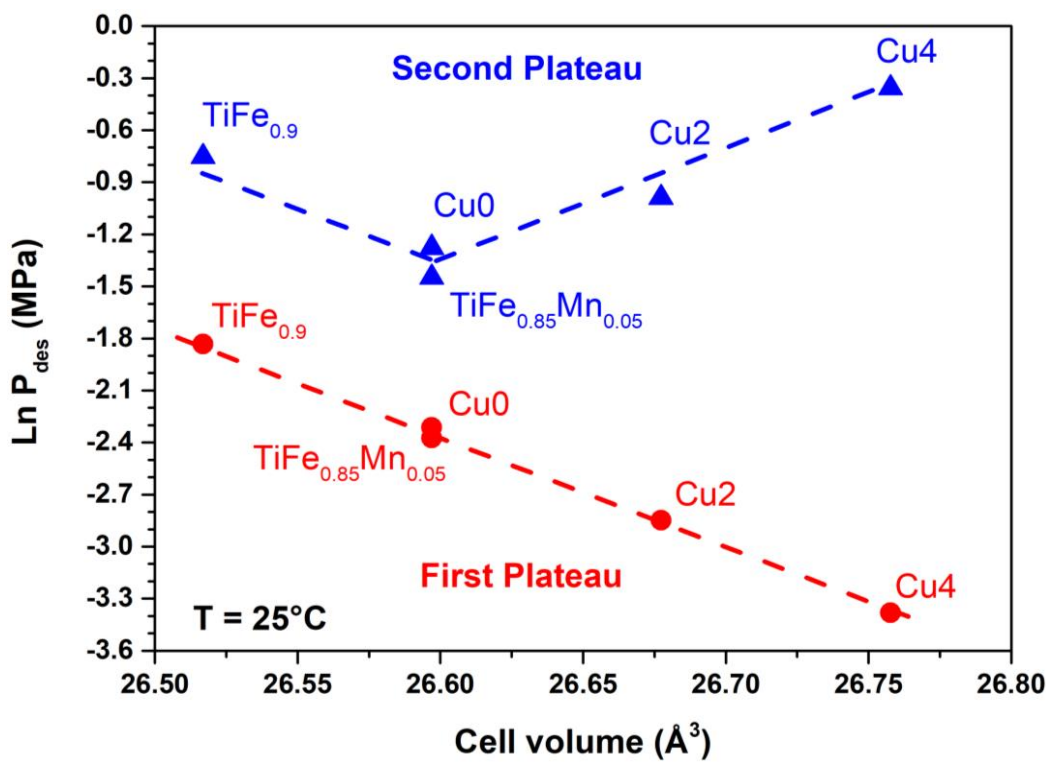

**Figure 4** – Linear dependences at 25 °C of the first (circles, red colour) and second (triangles, blue colour) plateau pressures on desorption as a function of the intermetallic cell volume for Cu0, Cu2 and C4 samples as well as literature data for TiFe$_{0.90}$[9] and TiFe$_{0.85}$Mn$_{0.05}$[10]. Dashed lines are guides to the eye.

Modification in plateau pressures with Cu substitution is reflected by the evolution of the enthalpy and entropy of hydrogen sorption which has been evaluated from Van't Hoff plots (**Table 4** and **Figure S5**). For the first plateau, the reaction enthalpy globally increases, in absolute values, both in absorption and desorption, which confirms that the monohydride is stabilized by Cu addition. In contrast, for the second plateau, enthalpy values tend to decrease with Cu content and the dihydride is destabilized. Note that this general observation is apparently not obeyed for Cu2. This is attributed to its higher entropy values (compared to Cu0 and Cu4) making the reaction enthalpy likely over-evaluated.

It is clearly established that the stoichiometric TiFe compound presents two distinct pressure plateaus.[4,5] They are observed as well for the Ti-rich off-stoichiometric TiFe$_{0.90}$ compound.[9]

However, the pressure difference between these two plateaus can be modified and tailored by elemental substitution. For instance, compared to TiFe$_{0.90}$,[9] Mn substitution reduces the pressure difference between the first and second plateau as it can be observed in this work for TiFe$_{0.88}$Mn$_{0.02}$, similarly to TiFe$_{0.85}$Mn$_{0.05}$[10] (**Figure S4**). While an almost single-plateau PCI curve is recorded for TiFe$_{0.88}$Mn$_{0.02}$ (**Figure 3, a**), the pressure difference between the two plateaus increases by Cu substitution, and it is especially high for TiFe$_{0.84}$Mn$_{0.02}$Cu$_{0.04}$ (**Figure 3, c**). Challet *et al*. studied the influence of Mn and Ni substitution for Fe in TiFe$_{0.90}$.[10] Their study shows similar substitutional effects as those here observed. They report that with Mn substitution, pressure difference between first and second plateau is increased at high temperature, making the pressure adjustment in a narrow pressure range difficult. While, when substituting with Ni, the pressure difference is increased even at low temperature in the same way as Cu does in this study.

Thermodynamic results presented in this work for Cu-substitution as well as those reported by Challet *et al.* for Ni,[10] clearly evidence that hydride stability of TiFe-based alloys cannot be solely predicted by geometric models. Cu substitution increases TiFe cell-volume but Ni-one has no significant effect. However, in both cases, the monohydride is stabilized whereas the dihydride is destabilized. This corroborates that electronic rather than geometric effects are at the basis of hydride stability in TiFe-compounds and more generally in Ti-based *AB* intermetallics with CsCl-type structure. As a matter of fact, chemical bonds between Ti and *B*-type elements depend strongly on the nature of late transition metals and affects hydrogen storage capacities and phase stability.[31–33] The extremely different hydrogenation properties of the end Ti*B* compounds TiFe, TiCo and TiNi with *B*-atoms having very close atomic radii (r = 1.24, 1.25 and 1.25 Å for *B* = Fe, Co and Ni, respectively) demonstrate this statement. TiFe absorbs up to 2 hydrogen atoms by formula unit, H/f.u., with two distinct plateau pressures. TiCo absorbs 1.5 H/f.u. with three different plateaux pressures.[34,35] Finally, hydrogen absorption in TiNi is limited to 1.4 H/f.u through a sloping plateau pressure.[34,35] The present work is an additional proof that the hydride stability of Ti-based *AB* intermetallics are governed by electronic properties and that empirical geometric rules are only an indirect consequence of electronic effects.

## Conclusions and outlook

Mn and Cu have been studied as elemental substituents in TiFe-based alloys to adapt the plateau pressure of the corresponding hydrides. All synthetized alloys present chemical homogeneity and flat PCI curves for the first pressure plateau thanks to the annealing treatment. They are characterised by a major TiFe matrix and minor contents of β-Ti solid solution and $Ti_4Fe_2O$ oxide precipitate phases. Low Mn amount ($TiFe_{0.88}Mn_{0.02}$) slightly modifies the hydrogen storage properties of $TiFe_{0.90}$. On the other hand, Cu increases the lattice parameter of the cubic TiFe-type phase efficiently decreasing the first plateau pressure. Augmenting Cu content, the amount of β-Ti secondary phase is increased, reducing the reversible capacity of the material and requiring harsher condition for activation. The reduced maximum capacity is also due to the increased pressure difference between the first and second plateau pressure of the intermetallic compound induced by Cu substitution. Substitutional effects on hydride stability cannot be only explained in terms of geometrical considerations. Electronic properties play a key role in the hydride stability of Ti-based *AB* intermetallics.

In conclusion, this work demonstrates that Mn and Cu substitutions in $TiFe_{0.9}$ intermetallic compound have significant effects on its hydrogen sorption properties that are relevant for the use of this compound in large-scale hydrogen storage at low pressure. Mn substitution lowers both plateau pressures in the TiFe-H system, whereas Cu one lowers the first plateau pressure while increasing the second one. Thus, equilibrium plateau pressures of TiFe(Mn,Cu) system can be finely adapted to the final application by careful variation of element substitution.

## Acknowledgement

This project is part of the HyCARE project, which has received funding from the Fuel Cells and Hydrogen 2 Joint Undertaking (JU) under grant agreement No 826352. The JU receives support from the European Union's Horizon 2020 research and innovation programme and Hydrogen Europe and Hydrogen Europe Research.

The authors wish to thank E. Leroy for EPMA analysis and F. Couturas for his help with hydrogenation experiments.

**Electronic Supporting Information Description**

ESI includes for all investigated sample a) Cu0, b) Cu2, c) Cu4: Rietveld refinements (before and after PCI), curves of pressure as a function of time with details on activation and kinetics, a comparison of PCI curves at 25°C of investigated samples and literature data, Van't Hoff plots used for the determination of thermodynamics.

# ESI

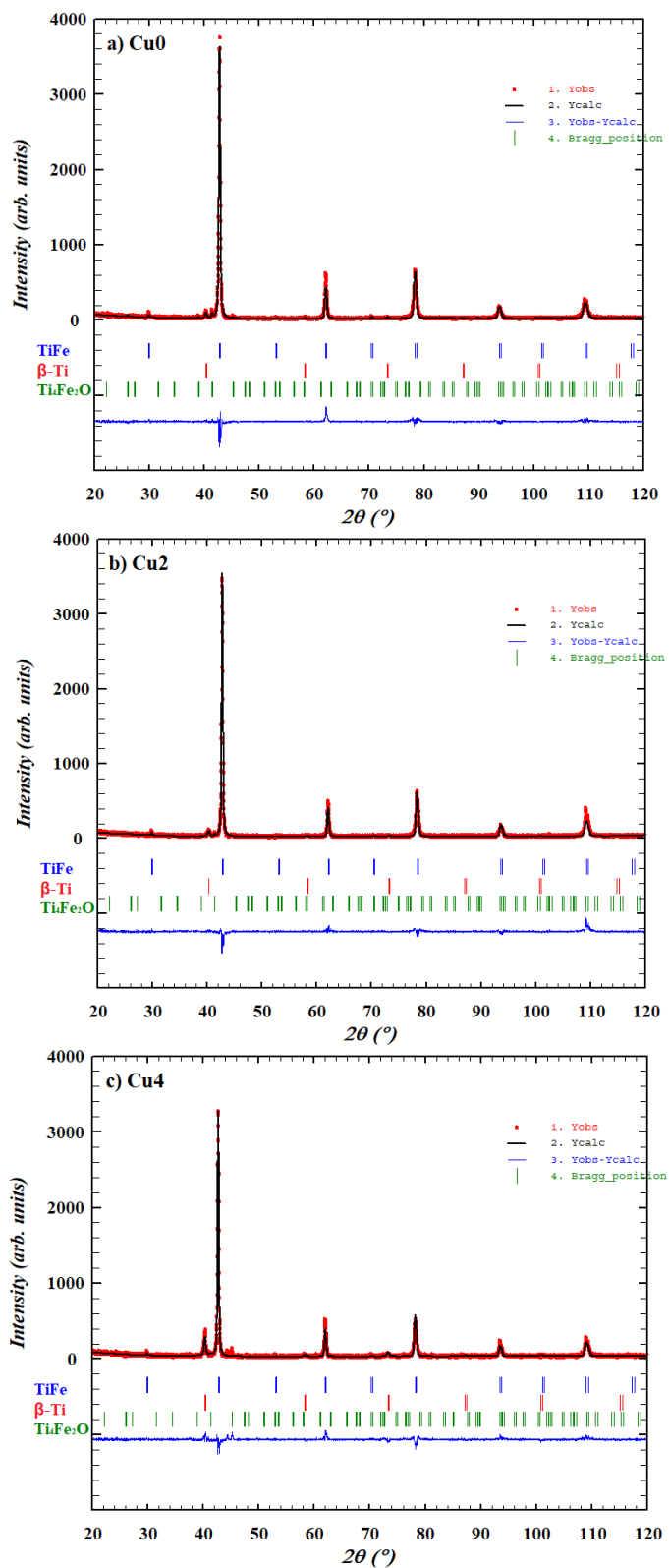

**Figure S1** – Rietveld refinements for investigated sample after annealing: a) Cu0, b) Cu2, c) Cu4

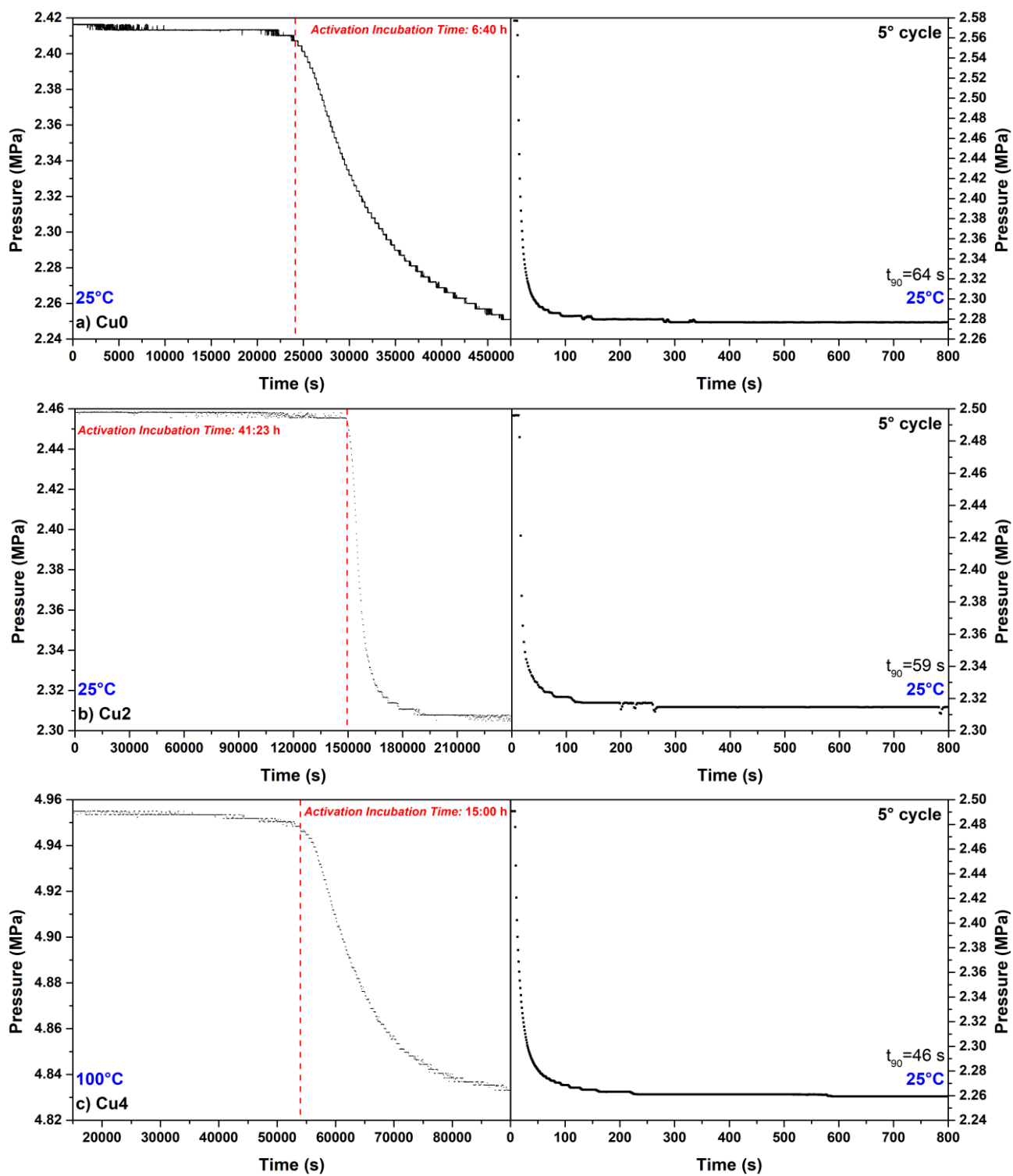

**Figure S2** – Pressure data as a function of time during first hydrogenation (activation) and during the fifth absorption after initial cycling, representative of the kinetic of the samples: a) Cu0, b) Cu2 and c) Cu4. Temperature, activation incubation time, and time to reach 90% of reacted fraction, $t_{90}$, is displayed as well.

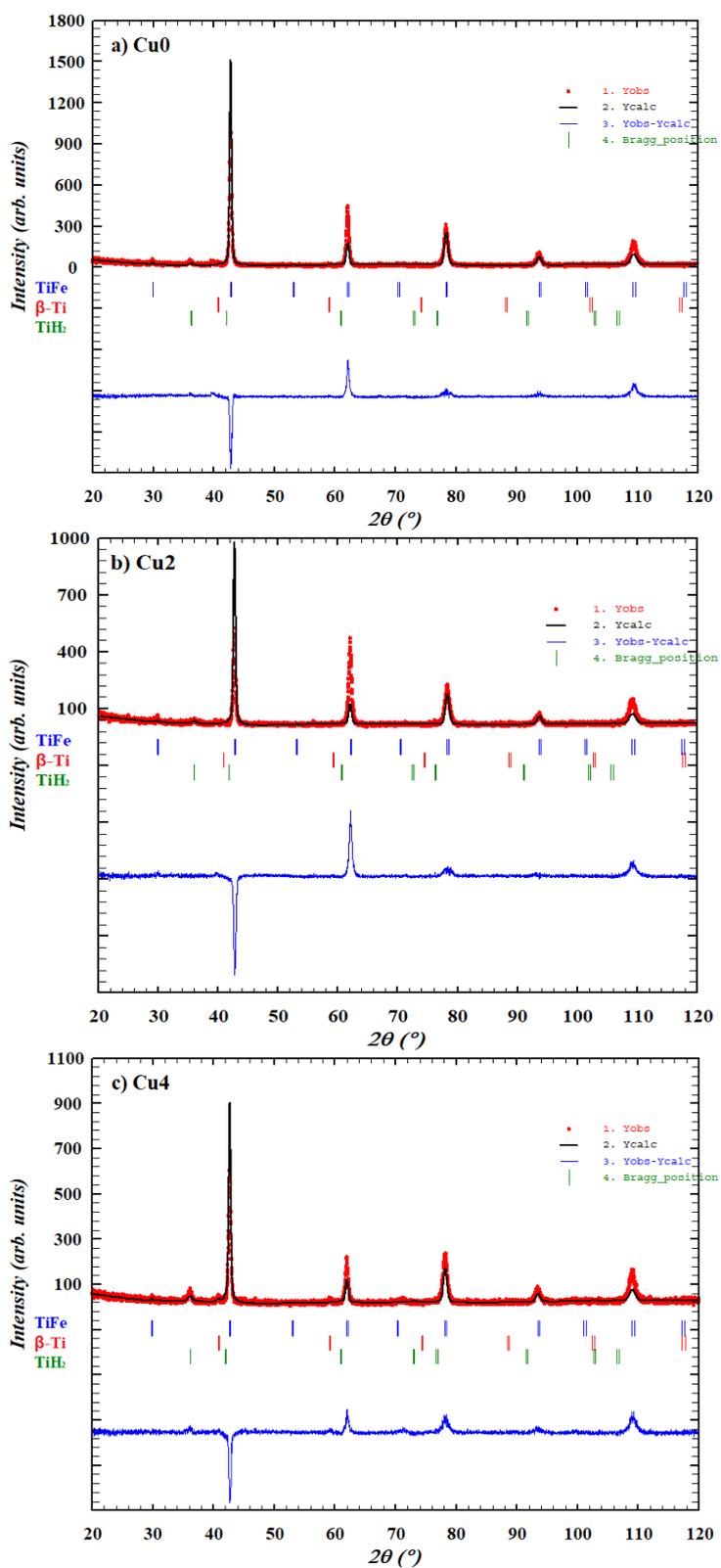

**Figure S3** – Rietveld refinements for investigated sample after PCI curves: a) Cu0, b) Cu2, c) Cu4

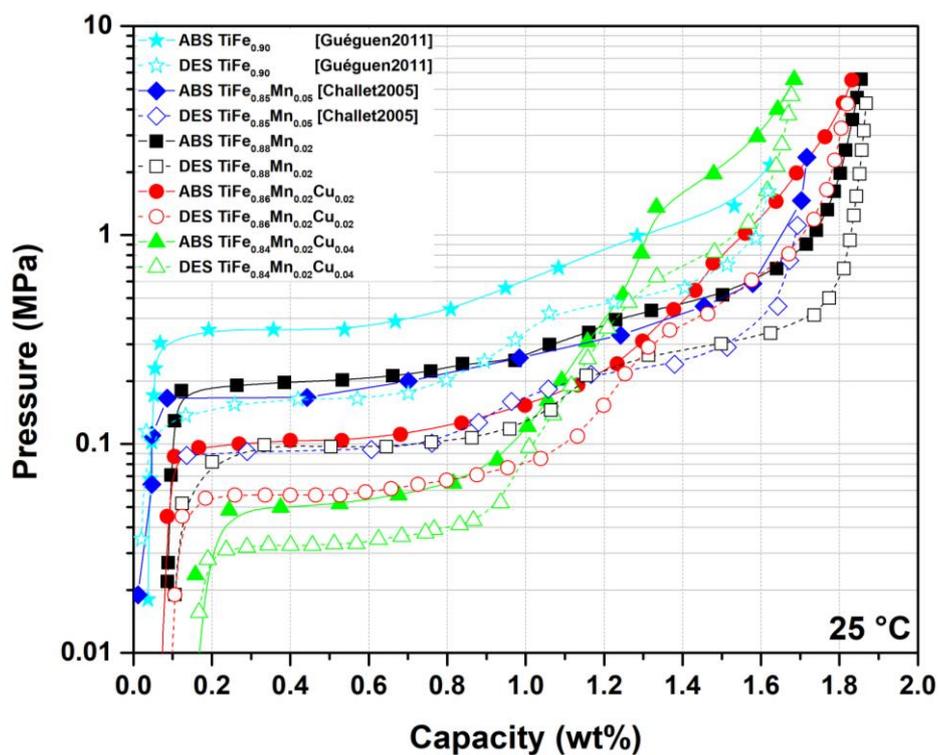

**Figure S4** – Comparison of PCI curves at 25°C upon hydrogenation (closed points) and desorption (open points) for Cu0 (TiFe$_{0.88}$Mn$_{0.02}$), Cu2 (TiFe$_{0.86}$Mn$_{0.02}$Cu$_{0.02}$) and Cu4 (TiFe$_{0.84}$Mn$_{0.02}$Cu$_{0.04}$) samples of this work and literature curves for TiFe$_{0.90}$ from [Guéguen2011][9] and TiFe$_{0.85}$Mn$_{0.05}$ from [Challet2005][10]. Lines in absorption and dashed lines in desorption are a guide for the eyes.

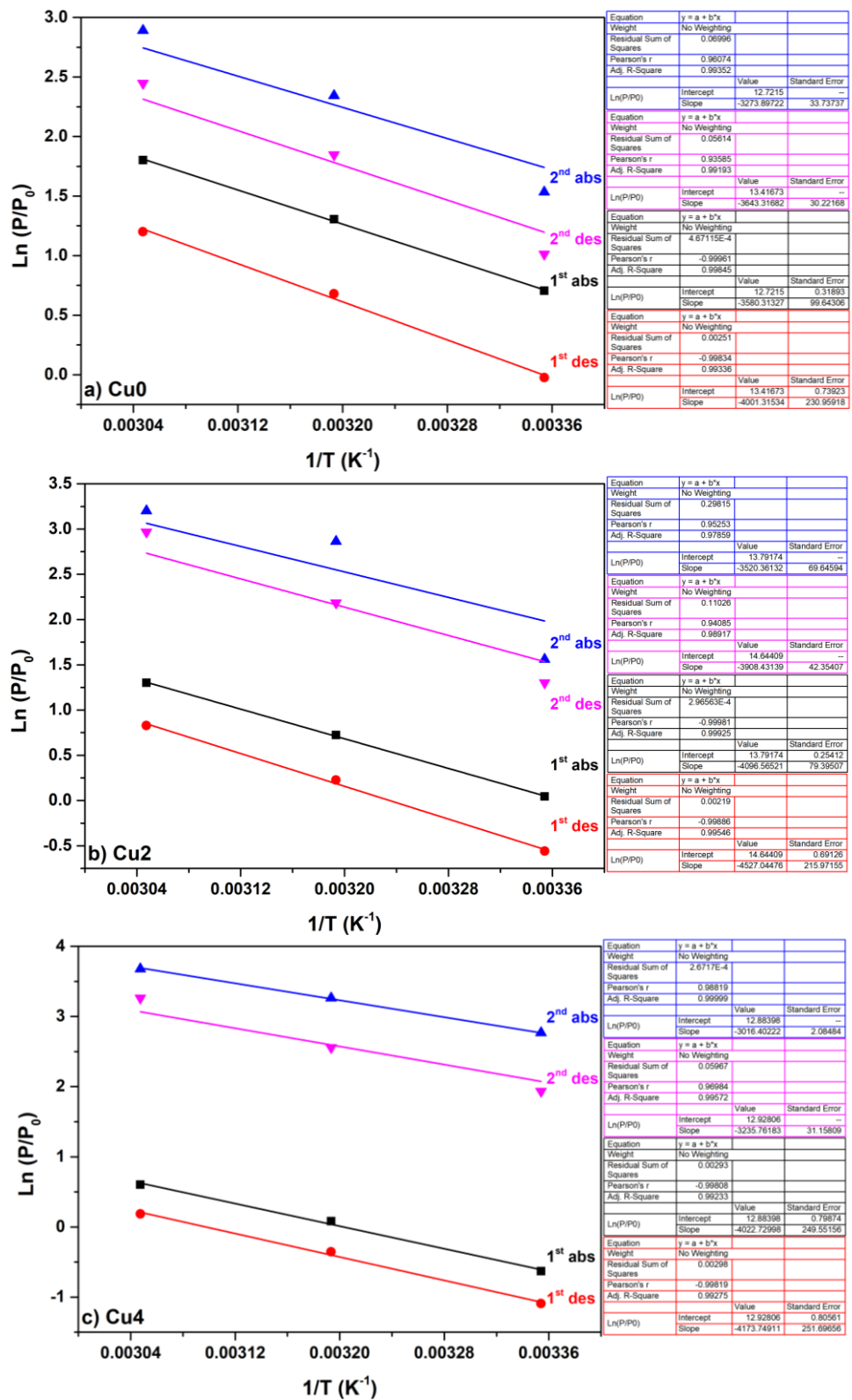

**Figure S5** – Van't Hoff plots for investigated samples: a) Cu0, b) Cu2, c) Cu4.

**Table S1** – Phase abundances and lattice parameter of TiFe phase for the three alloys here synthetized as determined by Rietveld analysis of XRD data after PCI curves.

| Sample | Nominal composition | TiFe a (Å) | TiFe wt.% | β-Ti wt.% | $TiH_2$ wt.% |
|---|---|---|---|---|---|
| **Cu0** | $TiFe_{0.88}Mn_{0.02}$ | 2.986(8) | 95.7±1.4 | 0.6±0.5 | 3.7±0.5 |
| **Cu2** | $TiFe_{0.86}Mn_{0.02}Cu_{0.02}$ | 2.989(7) | 93.2±1.9 | 0.9±0.5 | 5.9±0.5 |
| **Cu4** | $TiFe_{0.84}Mn_{0.02}Cu_{0.04}$ | 2.993(3) | 88.8±1.4 | 2.0±0.5 | 9.3±0.7 |